\documentclass[aps,preprintnumbers,epsf,amssymb,graphicx]{revtex4}
\usepackage[T1]{fontenc}           
\usepackage[latin1]{inputenc}      

\begin{document}
\title{Gauge invariant metric fluctuations in the early universe
from STM theory of gravity: nonperturbative formalism}
\author{$^{1,2}$Mariano Anabitarte \footnote{
E-mail address: anabitar@mdp.edu.ar}, $^{1,2}$Mauricio
Bellini\footnote{E-mail address: mbellini@mdp.edu.ar}}

\address{$^1$ Departamento de
F\'{\i}sica, Facultad de Ciencias Exactas y Naturales, Universidad
Nacional de Mar del Plata, Funes 3350, (7600) Mar del Plata,
Argentina.\\
$^2$ Consejo Nacional de Investigaciones Cient\'{\i}ficas y
T\'ecnicas (CONICET).}

\begin{abstract}
We develop a nonperturbative quantum field formalism to describe
scalar gauge-invariant metric flucturations in the early universe
from a 5D apparent (Ricci flat) vacuum.
\end{abstract}
\maketitle

\section{Introduction and Formalism}

The theory of linearized gravitational perturbations is a very
important issue of study in modern cosmology\cite{bardeen,riotto}.
It is used to describe the growth of structure in the universe, to
calculate the predicted microwave background fluctuations, and in
many other considerations. The growth of gravitational
perturbations in the universe is consequence of gravitational
instability. Consequently the perturbation will increase and will
in turn produce a more attractive force. In an expanding universe
this attraction is counteracted by the expansion. This, in
general, results in a power-law growth of the
perturbations\cite{AB}. The linearized approximation of these
perturbations are a very good description on cosmological scales,
because they are very small. However, to describe (scalar) gauge
invariant metric fluctuations on astrophysical scales, one should
consider a more general nonlinear (nonperturbative) study for
scalar metric fluctuations.

In particular, in this letter we shall consider a 5D formalism
developed on an apparent vacuum. We shall consider that the extra
dimension (which is a space like dimension) is noncompact. This
are typical ideas of the Space Time Matter (STM) theory of
gravity\cite{libro}, developed by Wesson and
co-workers\cite{Wesson1,Wesson2,Wesson3}. The STM theory of
gravity is based on an unrestricted 5D manifold, where the extra
dimension and derivatives with respect to the extra coordinate are
used to explain the origin of 4D matter in terms of geometry. This
goal was exposed by Einstein, who wished to transpose the
``base-wood'' of the marble of the left-hand side. That is, he
wished to find an algebraic expression for what is usually called
the energy-momentum tensor $T_{\alpha\beta}$, which nowadays refer
to as the Einstein tensor $G_{\alpha\beta}$ ($\alpha$ and $\beta$
run from $0$ to $3$). In other words, in STM theory of gravity one
assumes that the effective 4D Einstein equations $G_{\alpha\beta}
= -8\pi G T_{\alpha\beta}$ are induced from geometry from a 5D
apparent vacuum space on which $R_{AB}=0$ ($A$ and $B$ run from
$0$ to $4$). In particular, in this letter we shall develop a
quantum field formalism for scalar gauge invariant metric
fluctuations, where the fields of the physical system describe a
noncommutative algebra.

\subsection{Geometrical 5D apparent vacuum}

We consider the 5D background Riemann flat metric \cite{LB}
\begin{equation}\label{bm}
dS^2_b=l^2 dN^2-l^2 e^{2N} dr^2 -dl^2,
\end{equation}
which is 3D spatially isotropic, homogeneous and Riemann flat:
$R^A_{BCD}=0$. The metric (\ref{bm}) is a special case of the
much-studied class of canonical metrics $dS^2 = l^2 g_{\mu\nu}
dx^{\mu} dx^{\nu} - dl^2$\cite{can}. Furthermore, $N$, $x$, $y$,
$z$ are dimensionless coordinates and $l$, which is not considered
as compact, has spatial units. The issue of scalar gauge-invariant
metric fluctuations was recently studied in previous
works\cite{metric}, where these fluctuations were considered to be
small. In these works was used the perturbed metric $dS^2=l^2
(1+2\psi) dN^2- l^2(1-2\psi) e^{2N} dr^2 - (1-2\psi) dl^2$, where
$\psi(N,\vec r,l)$ are the gauge invariant (quantum) metric
fluctuations, which are considered to be very small. Hence, that
formalism should be, in principle, only valid to describe these
fluctuations in the early universe, and on cosmological scales.
With the aim to extend the validity of such theory to shorter
scales, we shall propose an extended perturbed metric
\begin{equation}\label{me1}
dS^2 =  l^2 e^{2\psi} dN^2 - l^2 e^{2(N- \psi)}dr^2-e^{-2\psi}
dl^2.
\end{equation}
Notice that in absence of fluctuations (i.e., when $\psi=0$), the
metric (\ref{me1}) gives us the 5D Riemann flat background metric
(\ref{bm}). We shall consider that the metric (\ref{me1})
describes an apparent 5D Ricci flat vacuum state: $R_{AB}=0$.
Furthermore, $dr^2=dx^2+dy^2+dz^2$, and $l$ is the noncompact
extra dimension. To describe the system we shall consider the
action
\begin{equation}
^{(5)}I = {\Large\int} d^4 x \  d\psi \sqrt{\left|\frac{^{(5)}
 g}{^{(5)} g_0}\right|} \left(
\frac{^{(5)} R}{16\pi G}+ \frac{1}{2}g^{AB} \varphi_{,A}
\varphi_{,B} \right),
\end{equation}
where $^{(5)}g$ is the determinant of the covariant metric tensor
$g_{AB}$ ($A$ and $B$ run from $0$ to $4$):
$^{(5)}g=l^8e^{6(N-\psi)}=(l^4e^{3(N-\psi)})^2$. Furthermore, we
shall consider a null scalar curvature of the perturbed metric
(\ref{me1}): $^{(5)} R=0$
\begin{eqnarray}\label{ricci}
^{(5)} R & = & \frac{4e^{2(\psi-N)}}{l^2}
\left\{\nabla^2\psi-(\nabla\psi)^2+ e^{2N} \left[ e^{-4\psi}
\left(-2\psi_{,NN} + 7 (\psi_{,N})^2 - 9
\psi_{,N} + 3\right) \right. \right.\nonumber \\
&& \left. \left.+ l^2 \left( \psi_{,ll}- (\psi_{,l})^2 \right) +3
l \psi_{,l}-3 \right] \right\}.
\end{eqnarray}
The Lagrange equations give us the equation of motion for the
fields $\psi$ and $\varphi$
\begin{eqnarray} \left( \frac{\partial^{(5)}R}{\partial\psi} - 3
\,{^{(5)} R} \right) & - & \left[ 3
\frac{\partial^{(5)}R}{\partial \psi_{,N}} + \frac{4}{l}
\frac{\partial^{(5)}R}{\partial \psi_{,l}} +
\frac{\partial}{\partial x^{A}} \left(
\frac{\partial^{(5)}R}{\partial \psi_{,A}}\right) \right]
\nonumber \\
& = & 4\pi G \left[ 5 l^{-2} e^{-2 \psi} (\varphi_{,N})^2 - l^{-2}
e^{2(\psi-N)} (\nabla \varphi)^2 - e^{2\psi} (\varphi_{,l})^2
\right], \label{mov1} \\
\varphi_{,NN}  +  \left( 3-5\psi_{,N}\right) \varphi_{,N}& - &
e^{4\psi-2N} \left(
\nabla^2\varphi-\vec\nabla\psi.\vec\nabla\varphi\right)
-l^2e^{4\psi}\left[\varphi_{,ll}
+\left(\frac{4}{l}-\psi_{,l}\right)\varphi_{,l}\right]=0.
\label{mov2}
\end{eqnarray}

\subsection{Energy Momentum tensor on an apparent 5D vacuum}

The 5D Energy-Momentum tensor for a scalar field $\varphi$ on the
metric (\ref{me1}), is
\begin{equation}
T_{AB} = \varphi_{,A} \varphi_{,B} - \frac{1}{2} g_{AB}
\varphi_{,C} \varphi^{,C},
\end{equation}
which is null, because we are considering an apparent vacuum on
the perturbed metric (\ref{me1}).The diagonal components are
\begin{eqnarray}
T_{NN} & = & \frac{1}{2} (\varphi_{,N})^2 + \frac{1}{2}
e^{4\psi-2N} (\nabla \varphi)^2 +\frac{l^2}{2} e^{4\psi}
(\varphi_{,l})^2, \label{t1} \\
T_{ll} & = &  \frac{1}{2} (\varphi_{,l})^2 + \frac{l^{-2}}{2}
e^{-4\psi} (\varphi_{,N})^2 - \frac{l^{-2}}{2} e^{-2N} (\nabla
\varphi)^2, \label{t2} \\
T_{xx} & = &  (\varphi_{,x})^2 + \frac{1}{2} e^{2N} e^{-4\psi}
(\varphi_{,N})^2 - \frac{1}{2} (\nabla \varphi)^2 - \frac{1}{2}
l^2 e^{2N} (\varphi_{,l})^2, \label{t3} \\
T_{yy} & = & (\varphi_{,y})^2 + \frac{1}{2} e^{2N} e^{-4\psi}
(\varphi_{,N})^2 - \frac{1}{2} (\nabla \varphi)^2 - \frac{1}{2}
l^2 e^{2N} (\varphi_{,l})^2, \label{t4} \\
T_{zz} & = & (\varphi_{,z})^2 + \frac{1}{2} e^{2N} e^{-4\psi}
(\varphi_{,N})^2 - \frac{1}{2} (\nabla \varphi)^2 - \frac{1}{2}
l^2 e^{2N} (\varphi_{,l})^2. \label{t5}
\end{eqnarray}
We can make the iddentification $T_{rr} = T_{xx} + T_{yy} +
T_{zz}$, and we obtain
\begin{equation}\label{tt}
T_{rr}  = \frac{3}{2} e^{2N} e^{-4\psi} (\varphi_{,N})^2 -
\frac{1}{2} (\nabla \varphi)^2  - \frac{3}{2} l^2 e^{2N}
(\varphi_{,l})^2.
\end{equation}

\subsection{Einstein equations on an apparent 5D vacuum}

The diagonal components (we consider
$G_{rr}=G_{xx}+G_{yy}+G_{zz}$) are all null, because, as well as
$T_{AB}=0$, we require that $G_{AB}=0$ in order to the vacuum on
the perturbed metric to be apparent. In other words, we are
considering that the system is in a true vacuum ($R^A_{BCD}=0$) on
the background metric (\ref{bm}), but in a apparent vacuum
($R_{AB}=0$ and then $G_{AB}=0$) on the perturbed metric
(\ref{me1}). Therefore
\begin{eqnarray}
G_{NN} =&& - 3 \left\{ e^{-2N+4\psi} \left[ \nabla^2 \psi -
(\nabla \psi)^2 \right] - 3 \frac{\partial \psi}{\partial N} +
2\left( \frac{\partial \psi}{\partial N} \right)^2  \right.\nonumber \\
&&\left.+ e^{4\psi} \left[ 3 l \frac{\partial \psi}{\partial l}
+l^2 \frac{\partial^2 \psi}{\partial l^2} - l^2 \left(
\frac{\partial \psi}{\partial l} \right)^2 \right] - \left[
e^{4\psi} -1 \right] \right\},
\end{eqnarray}
\begin{eqnarray}
G_{rr} =&& - 3 \left\{ e^{2N-4\psi} \left[ 10 \frac{\partial
\psi}{\partial N} - 9 \left( \frac{\partial \psi}{\partial N}
\right)^2 + 3 \frac{\partial^2 \psi}{\partial N^2} \right] +
e^{2N} \left[ -l \frac{\partial \psi}{\partial l} - l^2
\frac{\partial^2 \psi}{\partial l^2} + l^2 \left( \frac{\partial
\psi}{\partial l} \right)^2 \right] \right.\nonumber \\
&& \left.+ \frac{2}{3} \left[(\nabla \psi)^2 - \nabla^2 \psi
\right] + 3 e^{2N} \left[ 1 - e^{-4\psi} \right] \right\},
\end{eqnarray}
\begin{eqnarray}
G_{ll} = &&- \frac{1}{l^2} \left\{ e^{-2N} \left[(\nabla \psi)^2 -
\nabla^2 \psi \right] + e^{-4\psi} \left[ 15 \frac{\partial
\psi}{\partial N} - 9 \left( \frac{\partial \psi}{\partial N}
\right)^2 + 3 \frac{\partial^2 \psi}{\partial N^2} \right] \right.
\nonumber \\
&& \left. - 6 l \frac{\partial \psi}{\partial l} + 6 \left[ 1 -
e^{-4\psi} \right] \right\}.
\end{eqnarray}

Since we require that the Ricci scalar (\ref{ricci}) to be null:
$^{(5)} R=0$, we obtain
\begin{eqnarray}
\left[ e^{4\psi}-1 \right]=&& -\frac{1}{3} \left\{ 2
\frac{\partial^2 \psi}{\partial N^2} + 9 \frac{\partial
\psi}{\partial N}-7 \left(\frac{\partial \psi}{\partial
N}\right)^2 +  e^{4\psi-2N} \left[ (\nabla \psi)^2 -\nabla^2\psi \right] \right.  \nonumber \\
&&\left.+ e^{4\psi} \left[l^2 \left(\frac{\partial \psi}{\partial
l}\right)^2-3l \frac{\partial \psi}{\partial l} - l^2
\frac{\partial^2 \psi}{\partial l^2} \right]\right\}.\label{id}
\end{eqnarray}
Using the expression (\ref{id}) in the Einstein tensor components,
we obtain
\begin{eqnarray}
G_{NN} =&& - 2 \left\{ e^{4\psi-2N} \left[ \nabla^2 \psi - (\nabla
\psi)^2 \right] - \frac{1}{2}\left( \frac{\partial \psi}{\partial N} \right)^2
+\frac{\partial^2 \psi}{\partial l^2} \right.\nonumber \\
&&\left.+ e^{4\psi} \left[ 3 l \frac{\partial \psi}{\partial l}
+l^2 \frac{\partial^2 \psi}{\partial l^2} - 2 l^2 \left(
\frac{\partial \psi}{\partial l} \right)^2 \right]\right\}=0, \\
G_{rr} =&& - 3 \left\{ e^{2N-4\psi} \left[ \frac{\partial
\psi}{\partial N} - 2 \left( \frac{\partial \psi}{\partial N}
\right)^2 + \frac{\partial^2 \psi}{\partial N^2} \right] -
\frac{1}{3} \left[(\nabla \psi)^2 - \nabla^2 \psi \right] \right.\nonumber \\
&& \left. + e^{2N} \left[ -4l \frac{\partial \psi}{\partial l} -
2l^2 \frac{\partial^2 \psi}{\partial l^2} + 2l^2 \left(
\frac{\partial \psi}{\partial l} \right)^2 \right] \right\}=0,\\
G_{ll} = &&- \frac{1}{l^2} \left\{ -2 e^{-2N} \left[(\nabla
\psi)^2 - \nabla^2 \psi \right] + e^{-4\psi} \left[ -12
\frac{\partial \psi}{\partial N} + 12 \left( \frac{\partial
\psi}{\partial N} \right)^2 - 3 \frac{\partial^2 \psi}{\partial
N^2} \right] \right.
\nonumber \\
&& \left. + 3 l \frac{\partial \psi}{\partial l} - 3 l^2 \left(
\frac{\partial \psi}{\partial l} \right)^2 - 3 l^2
\frac{\partial^2 \psi}{\partial l^2}\right\}=0.
\end{eqnarray}
These are the effective 5D Einstein equations which will be used
to induce physics on an effective 4D spacetime.

\section{Effective 4D dynamics}

Now we consider the original 5D Riemann flat background metric
(\ref{bm}). We can make the transformation $N=H t$ and $l = l_0$,
so that the effective 4D metric for observers with velocities $U^l
=0$ will be
\begin{equation}
dS^2_b \rightarrow ds^2_b = l^2_0 H^2 dt^2 - l^2_0 e^{2H t} dr^2,
\end{equation}
which, once we take the foliation $l_0=1/H$, holds
\begin{equation}\label{b4}
ds^2_b = dt^2 - H^{-2} e^{2H t} dr^2.
\end{equation}
Here, $t$ is the cosmic time and $\vec r=|\vec r(x,y,z)|$ is
dimensionless. The metric (\ref{b4}) describes a background vacuum
dominated de Sitter like expansion: ${\rm p_b} =-\rho_b=-3H^2$,
where ${\rm p_b}$ is the background pressure and $\rho_b$ is the
background energy density. Notice that the source of the expansion
is induced by the foliation on the fifth coordinate: $\rho_b = 3
H^2=3/l^2_0$. Hence, we can introduce the effective 4D perturbed
metric of (\ref{b4})
\begin{equation}\label{4p}
ds^2 = e^{2\psi} dt^2 - H^{-2} e^{2 \left(H t-\psi\right)} dr^2,
\end{equation}
where $\psi\equiv \psi(t,\vec r)$. The effective 4D action $^{(4)}
I = \int d^4x \, ^{(4)} L$ will be
\begin{equation}
^{(4)} I = {\Large\int} d^4 x \sqrt{\left|\frac{^{(4)}
 g}{^{(4)} g_0}\right|} \left( \frac{^{(4)}
R}{16\pi G}+ \frac{1}{2}g^{\mu\nu} \varphi_{,\mu} \varphi_{,\nu} +
V \right),
\end{equation}
where $^{(4)} L$ is the Lagrangian
\begin{equation}\label{lag}
^{(4)} L = \sqrt{\left|\frac{^{(4)}
 g}{^{(4)} g_0}\right|} \left( \frac{^{(4)} R}{16\pi G}+
\frac{1}{2}g^{\mu\nu} \varphi_{,\mu} \varphi_{,\nu} + V \right),
\end{equation}
$V$ is the effective 4D potential induced by the foliation $l=1/H$
\begin{equation}
V = -\left.\frac{1}{2} g^{ll}
\left(\frac{\partial\varphi}{\partial l}\right)^2\right|_{N=H
t,l=l_0=1/H},
\end{equation}
and $^{(4)} R$ is the effective 4D Ricci scalar, related to the
metric (\ref{4p})
\begin{equation}\label{4ricci}
^{(4)} R = 6 \left[ 3 \left(\dot\psi\right)^2 - 5 \dot\psi H + 2
H^2 - \ddot\psi\right] e^{-2\psi}.
\end{equation}
With the aim of developing the effective 4D dynamics of the fields
$\varphi$ and $\psi$ we shall write the effective Einstein and
Lagrangian equations. The diagonal Einstein equations are

$\left.G_{NN}\right|_{N=H t,l=1/H} =-8\pi G
\left.T_{NN}\right|_{N=H t,l=1/H}$:

\begin{eqnarray}
&&e^{4\psi-2H t} \left[ \nabla^2 \psi - (\nabla \psi)^2 \right] -
\frac{H^2}{2}\left( \frac{\partial \psi}{\partial N} \right)^2
+\frac{\partial^2 \psi}{\partial l^2} + e^{4\psi} \left.\left[
\frac{3}{H} \frac{\partial \psi}{\partial l} +\frac{1}{H^2}
\frac{\partial^2 \psi}{\partial l^2} - \frac{2}{H^2} \left(
\frac{\partial \psi}{\partial l} \right)^2 \right]\right|_{N=H
t,l=1/H} \nonumber
\\
&&=  2 \pi G \left.\left[H^2 \left( \frac{\partial
\varphi}{\partial t}\right)^2 + e^{4\psi-2H t} (\nabla \varphi)^2
+ \frac{1}{H^2} e^{4\psi} \left(\frac{\partial \varphi}{\partial
l}\right)^2 \right]\right|_{N=H t,l=1/H},
\end{eqnarray}

$\left.G_{rr}\right|_{N=H t,l=1/H}=-8\pi G
\left.T_{rr}\right|_{N=H t,l=1/H}$:

\begin{eqnarray}
&& e^{2H t-4\psi} \left[ H\frac{\partial \psi}{\partial t} - 2 H^2
\left( \frac{\partial \psi}{\partial t} \right)^2 + H^2
\frac{\partial^2 \psi}{\partial t^2} \right] - \frac{1}{3}
\left[(\nabla \psi)^2 - \nabla^2 \psi \right] - e^{2H t}
\left.\left[ \frac{4}{H} \frac{\partial \psi}{\partial l} +
\frac{2}{H^2} \frac{\partial^2 \psi}{\partial l^2} - \frac{2}{H^2}
\left( \frac{\partial \psi}{\partial l} \right)^2
\right]\right|_{N=H t,l=1/H}
\nonumber \\
&&= 4 \pi G \left.\left\{ H^2 e^{2H t} e^{-4\psi} \left(
\frac{\partial \psi}{\partial t} \right)^2 - \frac{1}{3} (\nabla
\varphi)^2  -
 \frac{1}{H^2} e^{2H t} \left(\frac{\partial \varphi}{\partial
l}\right)^2 \right\}\right|_{N=H t,l=1/H},
\end{eqnarray}

$\left.G_{ll}\right|_{N=H t,l=1/H}=-8\pi G
\left.T_{ll}\right|_{N=H t,l=1/H}$:

\begin{eqnarray}
&& e^{-4\psi} \left.\left[ 12 H^2  \left( \frac{\partial
\psi}{\partial t} \right)^2 -12 H \frac{\partial \psi}{\partial t}
- 3 H^2 \frac{\partial^2 \psi}{\partial t^2} \right]  -2 e^{-2 H
t} \left[(\nabla \psi)^2 - \nabla^2 \psi \right] + \frac{3}{H}
\frac{\partial \psi}{\partial l} - \frac{3}{H^2} \left(
\frac{\partial \psi}{\partial l} \right)^2 - \frac{3}{H^2}
\frac{\partial^2 \psi}{\partial l^2}\right|_{N=H t,l=1/H} \nonumber \\
&& = 4 \pi G \left.\left\{\frac{1}{H^2} \left(\frac{\partial
\varphi}{\partial l}\right)^2 +  H^2 e^{-4\psi}
\left(\frac{\partial \varphi}{\partial t}\right)^2 - e^{-2H t}
(\nabla \varphi)^2\right\}\right|_{N=H t,l=1/H},
\end{eqnarray}
where $T_{NN}$, $T_{rr}$ and $T_{ll}$ are given respectively by
eqs. (\ref{t1}), (\ref{tt}) and (\ref{t2}). On the other hand, the
effective 4D Lagrange equations become from taking the equations
(\ref{mov1}) and (\ref{mov2}) on the foliation $l=1/H$ and the
transformation $N=H t$
\begin{eqnarray}
\left( \frac{\partial^{(4)}R}{\partial\psi} - 3 \,{^{(4)} R}
\right) & - & \left.\left[ 3 H \frac{\partial ^{(4)}R}{\partial
\dot\psi} + 4 H \frac{\partial ^{(4)}R }{\partial \psi_{,l}} +
\frac{\partial}{\partial x^{\mu}} \left(
\frac{\partial^{(4)}R}{\partial \psi_{,\mu}}\right)+
\frac{\partial}{\partial l} \left( \frac{\partial^{(4)}R}{\partial
\psi_{,l}} \right) \right]\right|_{l=1/H}
\nonumber \\
& = & 4\pi G \left.\left[ 5  e^{-2 \psi} (\dot\varphi)^2 - H^2
e^{2(\psi-H t)} (\nabla \varphi)^2 - e^{2\psi} (\varphi_{,l})^2
\right]\right|_{l=1/H}, \label{mov11} \\
\ddot\varphi  +  \left( 3H -5\dot\psi\right) \dot\varphi& - &
e^{4\psi-2H t} \left(
\nabla^2\varphi-\vec\nabla\psi.\vec\nabla\varphi\right)
-\frac{e^{4\psi}}{H^2} \left.\left[\varphi_{,ll} +\left(4
H-\psi_{,l}\right)\varphi_{,l}\right]\right|_{l=1/H}=0.
\label{mov22}
\end{eqnarray}
Note that the term $-5\dot\psi \dot\varphi$ in the equation for
$\varphi$ (\ref{mov22}), can be related with local dissipation,
which appears in warm and fresh inflationary
models\cite{berera1,berera2,bellini}. Of course, the dynamics of
the effective 4D system developed in this section is nonlinear,
and thus unsolvable. A linear approach was studied in two recent
papers\cite{metric}.

Since the field $\varphi(t,\vec r)$ is quantum in origin, it
should describe the following algebra:
\begin{equation}
\left[\varphi(t,\vec r), \Pi_{\varphi}(t,\vec r')\right] = i \,
g^{tt} \sqrt{\left|\frac{^{(4)}
 g}{^{(4)} g_0}\right|}\,e^{-\int
\left[3H-5\dot\psi\right] dt} \,\delta^{(3)}\left(\vec r-\vec r'\right), \\
\end{equation}
where $^{(4)} g=\left(1/H^3 e^{3Ht} e^{-2\psi}\right)^2$ is the
determinant of the effective 4D perturbed metric tensor
$g_{\mu\nu}$. Furthermore, the momentum $\Pi_{\varphi}$ is given
by
\begin{equation}
\Pi_{\varphi} = \frac{\partial ^{(4)} L}{\partial\dot\varphi},
\end{equation}
where the effective 4D Lagrangian is given by (\ref{lag}).

\section{Final Comments}

In this letter we have proposed a nonperturbative formalism to
gauge-invariant scalar (quantum) metric fluctuations, which evolve
in the early universe. In the model, the expansion of the universe
is driven by a single (quantum) scalar field. We have considered
the system from a 5D perturbed Ricci flat metric, on which we
assume an apparent vacuum state. Hence, all 4D sources become from
the geometrical foliation on the 5D metric. Since the formalism
here considered is nonpertubative, it is valid for strong
fluctuations, which could be produced on very small sub-Planckian
scales. Of course, in the weak field limit: $e^{\pm 2\psi} \simeq
1\pm 2\psi$, and this formalism should be equivalent to the one
developed in\cite{metric}, which only is
valid on cosmological (super Hubble) scales.\\

\centerline{\bf{Acknowledgements}} \vskip .2cm MA and MB
acknowledge CONICET and UNMdP (Argentina) for financial support. \\

\end{document}